\documentstyle[aps,prb,preprint]{revtex}
\begin{document}
\draft
\author{N. F. Schwabe\cite{Permanent}, A. N. Cleland, M. C. Cross, M. L.
Roukes}
\title{Perturbation of Tunneling Processes by Mechanical Degrees of Freedom in
Mesoscopic Junctions}
\address{Condensed Matter Physics 114-36\\
California Institute of Technology, Pasadena, CA 91125}
\date{\today }
\maketitle

\begin{abstract}
We investigate the perturbation in the tunneling current caused by
non-adiabatic mechanical motion in a mesoscopic tunnel junction. A theory
introduced by Caroli et al. \cite{bi1,bi2,bi3} is used to evaluate second
order self-energy corrections for this non-equilibrium situation lacking
translational invariance. Inelastic signatures of the mechanical degrees of
freedom are found in the current-voltage $I(V)$ characteristics. These give
rise to sharp features in the derivative spectrum, $d^2I/dV^2$.
\end{abstract}
\vspace{3ex}
\pacs{PACS numbers: 73.40.Gk 72.10.Di 85.30.Mn 85.42.+m}

\section{Introduction}

Electron tunneling has provided valuable insight into the physical
properties of solid state systems. Measurement of the current through
stationary tunnel junctions has been used to elucidate the density of states
in superconductors and many-body properties of metals and semiconductors.
Through much theoretical effort, a satisfactory theoretical description of
the tunneling process in condensed matter has been attained. Recently,
microfabrication techniques have progressed to the point where it is now
possible to make extremely compliant vacuum tunneling electrodes which may
not remain mechanically stationary during the tunneling process. In the
following work we consider the effect that such motion within an elastic
tunneling barrier may have upon electron tunneling characteristics. Since,
to our knowledge, previous studies of this question have been
phenomenological \cite{m1,m2,m3}, we shall attempt to present a more
complete theoretical approach to this problem.

We investigate the characteristics of the tunneling current through a square
potential barrier where the barrier is fixed on one side and is allowed to
oscillate freely on the other. The goal of this work is thus to predict the
effect of a mechanically compliant electrode, which can recoil from a
tunneling process, on the current-voltage characteristics and its first and
second derivatives. Mathematically this results in the treatment of a
localized phonon representing the movable part of the barrier.

In order to treat this problem in a many body approach which can be extended
to more realistic situations than the simplified model considered here, we
use a theory treating translationally non-invariant systems under
non-equilibrium situations developed in a series of papers by Caroli,
Combescot, Nozi\`{e}res and Saint-James \cite{bi1,bi2,bi3} (hereafter
referred to as CCNS) based on the Keldysh non-equilibrium perturbation
formalism \cite{bi4}. The theory gives a rigorous derivation of an energy
dependent transfer term from first principles, thereby extending the first,
more phenomenological approach of Bardeen\cite{bi10}, and it allows for a
treatment of the phonon perturbation in the usual diagrammatic theory.

The principle introduced in Ref.\onlinecite{bi1} can be used to treat a
non-equilibrium situation such as occurs in metal insulator metal (M-I-M)
tunneling. It consists of making one or more partitions in the system,
allowing for a separate treatment of regions in equilibrium. The separate
parts are then joined through an appropriate transfer term. For an arbitrary
one-electron potential Caroli {\em et al}. showed that this method yields an
exact treatment. CCNS then derived in Ref.\onlinecite{bi2} the well-known
expression for the tunneling current $I(V)$ through a rigid square barrier
for a quasi-equilibrium situation. Even though the series of papers starts
with a discrete model, CCNS later make a transition to a continuous
representation which we use for the calculations in the present paper. They
then consider electron-phonon interaction effects in two following papers
\cite{bi3,bi6}, but the discussion of the effects remained incomplete since
a treatment of real phonons is rather involved. The reason is that, for many
body potentials, the evaluation has to be done more carefully, as
renormalizations due to many body effects will always depend on the whole
system rather than just pieces of it. However in special cases such as the
one presented here, it can be shown that a simple approximation will work
very well.

\section{Outline of the calculation}

Our physical system consists of a mechanically-compliant cantilevered metal
tip of mass $m_c$, placed a small distance $2a$ from a stationary bulk metal
counter-electrode (see Fig. 1). The movable tip assembly, which we shall
refer to as the ``cantilever'', is modeled as a spring with Hooke's law
force constant $k_c$. Electrical contacts are made to the tip and the
counter-electrode. A voltage $V$ is applied across these electrodes and the
resulting characteristic of the current $I$ is measured as a function of the
applied voltage. In our model Hamiltonian of the system, the metal tip is
considered to be a single-mode harmonic oscillator with characteristic
frequency $\omega _c$. Assuming a quasi equilibrium situation of equal Fermi
energies $(\epsilon _F^r\simeq \epsilon _F^l)$ and the simple case of equal
work functions for both sides, the resulting potential barrier can be
approximated by a square barrier with barrier height $V_0$. The oscillation
of the cantilever simply adds an additional degree of freedom, resulting in
a modulation of the barrier width on one side (see Fig. 2). Hence our
quantum mechanical Hamiltonian reads

\begin{equation}
H=\frac{\vec{p}_e^2}{2m_e}+\frac{p_c^2}{2m_c}+\frac 12k_cx_c^2+V_0[\theta
(x_e+a)-\theta (x_e+x_c-a)]\text{ }  \label{a1}
\end{equation}
where the subscripts $e$ and $c$ refer to the tunnelling electron and the
cantilever respectively, while $x_c$ is the instantaneous displacement of
the cantilever from its equilibrium position, and $p_c$ is the corresponding
momentum. Also $\theta $ is the usual unit step function.

To treat the additional term in perturbation theory the potential term is
expanded to first order in $x_c$, which yields $V_r(x_e,x_c)=V_0[\theta
(x_e+a)-\theta (x_e-a)]$ $-x_cV_0\delta (x_e-a)$. In order to apply a many
body treatment to this problem we follow the approach of Ref.\onlinecite{bi2}
and split the system into two halves by adding an infinite potential barrier
in the middle of the unperturbed barrier. We thus obtain two sub-parts
similar to the appendix of Ref.\onlinecite{bi3} with the potential terms $%
V_r(x_e,x_c)=V_0\theta (-x_e+a)-x_cV_0\delta (x_e-a)$ for $x>0$ and $%
V_r(x_e,x_c)=\infty $, for $x<0$ for the right hand side medium and $%
V_l(x_e,x_c)=$ $V_0\theta (x_e+a)$ for $x<0$ and $V_l(x_e,x_c)=\infty $, for
$x>0$ for the left one (see Fig. 3). As in Ref.\onlinecite{bi3} we shall
assume that these two resulting sub-parts are semi-infinite in the direction
perpendicular to the barrier, while at the same time having a finite cross
sectional area $\nu ^{(2)}$ on which we impose periodic boundary conditions.
In second quantized notation the Hamiltonian for the right side reads
\begin{equation}
H_r=\sum_{\vec{q}}\frac{\vec{k}_{\vec{q}}^2}{2m_e}c_{\vec{q}}^{+}c_{\vec{q}%
}+\omega _c(a^{+}a+\frac 12)-x_0V_0\sum_{q_x,q_{x^{\prime }},q_{\parallel
}}c_{q_x,q_{\parallel }}^{+}c_{q_{x^{^{\prime }}},q_{\parallel }}\phi
_{q_x}^{*}(a)\phi _{q_{x^{^{\prime }}}}(a)(a^{+}+a)\text{ }  \label{a2}
\end{equation}
where the $x_0=\sqrt{1/2m_c\omega _c},$ the root mean square zero point
displacement of the cantilever, and we use $\hbar \equiv 1$. Also the
variable $x$ labels the coordinate perpendicular to the barrier. In this
representation the wave functions $\phi _{q_x}$ are the $x$-dependent
components of wave functions $\chi _{\vec{q}}$ which diagonalize the system
when the unperturbed barrier is included; we will discuss their explicit
form later. We will follow the procedure used in Ref.\onlinecite{bi2} to
obtain an expression for the current across the barrier $I(V),$ see their
Eq. (41), (46), (47), and for $dI/dV$ , $d^2I/dV^2$ as well as corrections
to these quantities, following the procedure in Ref.\onlinecite{bi3}. As
indicated in our model Hamiltonian, we assume that the barrier width does
not fluctuate spatially---this leads to perfectly specular transmission. The
unperturbed Green's functions for both parts of the system have been
calculated in the appendix of Ref.\onlinecite{bi3} and will be used as the
starting point for further calculations.

Following Ref.\onlinecite{bi3}, to calculate the current we first obtain a
self energy correction for the unperturbed Green's functions of our system,
which are then used to determine the corrections for the function $\gamma $
introduced in Ref.\onlinecite{bi2}, Eqs. (17), (30). (Also see below for
definitions.) The first order terms in the S-matrix expansion for the self
energy vanish, and so we consider the second order direct and exchange
diagrams (see Fig. 4). These terms are dominant due to the heavy mass of the
cantilever\cite{bi5}. A three dimensional calculation for the Green's
function is used to provide a more realistic situation than the one
dimensional model introduced by Caroli {\em et al}. Due to the translational
symmetry in the directions parallel to the barrier, we Fourier transform our
Green's functions in these directions, and we obtain the expressions for $%
I(V)$ by integrating over the distribution of possible $k_{\parallel }$%
-vectors. For the dispersion we assume a single parabolic band and leave the
electronic rest mass unaltered for simplicity. We adopt the notation used by
CCNS, that is:

\begin{description}
\item  $G$ --- total propagator of the coupled system including phonon
corrections;

\item  $G^0$ --- propagator of the total system without phonon corrections;

\item  $g$ --- propagator of the right hand side infinite medium including
phonon corrections;

\item  $g^0$ --- propagator of the right hand side infinite medium without
phonon corrections;

\item  $\tilde{g}$ --- propagator of the left hand side infinite medium
including phonon corrections;

\item  $\tilde{g}^0$ --- propagator of the left hand side infinite medium
without phonon corrections; and

\item  $\gamma (x_0)\equiv -\frac 1{2m}\lim_{x,x^{\prime }\rightarrow
x_0}\partial _x\partial _{x^{\prime }}g(x,x^{\prime };k_{\parallel };\omega
) $ (with $\tilde{\gamma}$ defined accordingly).
\end{description}

The quantity $\gamma $ is related to the Green's functions of the
semi-infinite media, which was introduced in Eq. (17) of Ref.\onlinecite{bi2}%
; we refer the reader there for a detailed treatment.

The corresponding expressions for $g^0$ and $\tilde{g}^0$ were derived in
the appendix of Ref.\onlinecite{bi3}. In our case the three dimensional
treatment alters the $\omega $ dependence slightly. The resulting
differential equation for the semi-infinite medium on the right hand side,
after Fourier-transformation in the $k_{\parallel }$ direction, now reads
\begin{equation}
\left( -\frac 1{2m}(\frac{\partial ^2}{\partial x^2}-k_{\parallel
}^2)+V_r(x)-(\omega +\mu )\right) g^{0r}(x,x^{\prime };k_{\parallel };\omega
)=\delta (x-x^{\prime })\text{ }  \label{a3}
\end{equation}
where the superscript $r$ denotes the retarded function and $V_r(x)$ has the
shape shown in Fig. 3.{\bf \ }

Explicitly, the unperturbed retarded Green's function $g^{0r}$ is given by
\begin{equation}
g^{0r}(x,x^{\prime };k_{\parallel };\omega )=\left\{
\begin{array}{ll}
-2m%
{\displaystyle {\sinh Kx \over KD}}
[(K+iq)\exp \{K(x^{\prime }-a)\}-(K-iq)\exp \{K(a-x^{\prime })\}]\text{ } &
x<x^{\prime }<a \\
-2m%
{\displaystyle {\sinh Kx^{\prime } \over KD}}
[(K+iq)\exp \{K(x-a)\}-(K-iq)\exp \{K(a-x)\}] & x^{\prime }<x<a \\
-2m%
{\displaystyle {\sinh Kx^{\prime } \over KD}}
2K\exp \{iq(x-a)\} & x>a
\end{array}
\right.  \label{a4}
\end{equation}
where $K=\sqrt{2m(V_0-(\omega +\mu -\frac{k_{\parallel }^2}{2m}))}$ and $q=%
\sqrt{2m(\omega +\mu -\frac{k_{\parallel }^2}{2m})}$ denote the wave vectors
inside and outside the barrier respectively, and the denominator $D$ is
given by $D=(K+iq)\exp (-Ka)+(K-iq)\exp (Ka)=2(K\cosh (Ka)-iq\sinh (Ka)).$
For small barrier transparencies, which are considered here, $\exp (Ka)$ $%
\gg 1$ and the denominator is then approximately $D=(K-iq)\exp Ka$.

\section{Self energy corrections}

\subsection{Dyson equation for the self energy}

Since our system is not translationally invariant we must resort to a Dyson
equation in coordinate representation for the direction perpendicular to the
barrier . We make the approximation, which will be justified in the
following, that only the electronic subsystem on the right will receive any
corrections from coupling to the mechanical degree of freedom. As the
calculations for the unperturbed Green's functions on both sides are
analogous, we will confine ourselves to calculating $g$, the propagator of
the right subsystem. Ultimately, as shown in Ref.\onlinecite{bi2}, it is
necessary to obtain corrections to the Green's functions $g^{0+}$ and $%
g^{0-} $occurring in the Keldysh formalism in order to calculate a
correction to the current. They are defined in Keldysh \cite{bi4} and also
in Eq. (15) of Ref.\onlinecite{bi2} as

\begin{eqnarray}
g^{0+}(\omega ) &=&2\pi i\rho ^0(\omega )f^0(\omega )  \label{b0} \\
g^{0-}(\omega ) &=&2\pi i\rho ^0(\omega )\{f^0(\omega )-1\}\qquad .
\nonumber
\end{eqnarray}
Here $\rho ^0(\omega )$ and $f^0(\omega )$ are the spectral density and the
Fermi occupation function of the right subsystem, respectively. However, in
the quasi equilibrium situation considered here, these can be obtained from
the renormalized retarded Green's function $g^r$. The corresponding Dyson
equation reads
\begin{equation}
g^r(x,x^{\prime };k_{\parallel };\omega )=g^{0r}(x,x^{\prime };k_{\parallel
};\omega )+\int dx^{\prime \prime }dx^{\prime \prime \prime
}g^{0r}(x,x^{\prime \prime };k_{\parallel };\omega )\tilde{\Sigma}%
^r(x^{\prime \prime },x^{\prime \prime \prime };k_{\parallel };\omega
)g^r(x^{\prime \prime \prime },x^{\prime };k_{\parallel };\omega )\quad .
\label{b1}
\end{equation}
Due to the localization of the electron-cantilever interaction the self
energy has the form
\begin{equation}
\tilde{\Sigma}^r(x^{\prime \prime },x^{\prime \prime \prime };k_{\parallel
};\omega )=\delta (x^{\prime \prime }-a)\delta (x^{\prime \prime \prime
}-a)\Sigma ^r(a,a;k_{\parallel };\omega )\quad  \label{b2}
\end{equation}
and the integration simplifies to
\begin{equation}
g^r(x,x^{\prime };k_{\parallel };\omega )=g^{0r}(x,x^{\prime };k_{\parallel
};\omega )+g^{0r}(x,a;k_{\parallel };\omega )\Sigma ^r(a,a;k_{\parallel
};\omega )g^r(a,x^{\prime };k_{\parallel };\omega )\text{ .}  \label{b3}
\end{equation}
Expressed in terms of the unperturbed Green's function this is
\begin{equation}
g^r(x,x^{\prime };k_{\parallel };\omega )=g^{0r}(x,x^{\prime };k_{\parallel
};\omega )+\frac{g^{0r}(x,a;k_{\parallel };\omega )\Sigma
^r(a,a;k_{\parallel };\omega )g^{0r}(a,x^{\prime };k_{\parallel };\omega )}{%
1-g^{0r}(a,a;k_{\parallel };\omega )\Sigma ^r(a,a;k_{\parallel };\omega )}%
\text{ .}  \label{b4}
\end{equation}

\subsection{The self energy in Migdal's approximation}

In our system, the cantilever mass $m_c$ is much larger than the electron
mass $m_e$ so that Migdal's approximation \cite{bi5} holds well. The self
energy corrections will be small and the largest contribution to the self
energy will come from the second order direct and exchange diagrams (see
Fig.4). Higher order diagrams will only contribute as $\sqrt{\frac{m_e}{m_c}}
$ , which in our case is certainly small. In view of the fact that we are
calculating the lowest relevant orders of the corrections, we also neglect
the second term in the denominator of (\ref{b4}). In the Matsubara
representation the direct and exchange diagrams are\\
\begin{equation}
\Sigma ^{1d}(a,a;k_{\parallel };i\omega _n)=-(x_0V_0)^2D^0(0)\frac{\nu ^{(2)}%
}\beta \int \frac{d^2k_{\parallel }^{\prime }}{(2\pi )^2}\sum_{i\omega
_n^{\prime }}G^0(a,a;k_{\parallel }^{\prime };i\omega _n^{\prime })\text{ }
\label{b5}
\end{equation}
and
\begin{equation}
\Sigma ^{1ex}(a,a;k_{\parallel };i\omega _n)=-(x_0V_0)^2\frac 1\beta
\sum_{i\omega _n^{\prime }}D^0(i\omega _n^{\prime })G^0(a,a;k_{\parallel
};i\omega _n-i\omega _n^{\prime })\quad .\text{ }  \label{b6}
\end{equation}
In (\ref{b5}) $\Sigma ^{1d}$ does not depend on $k_{\parallel }$, $\nu
^{(2)} $ is the cross sectional area of the junction and $\beta =(k_BT)^{-1}$%
. The single mode phonon propagator is defined as
\begin{equation}
D^0(i\omega _n)=\frac{2\omega _c}{\omega _c^2+\omega _n^2}\quad .  \label{b7}
\end{equation}
The fact that there is no integration over the parallel momentum in the
exchange term results from the fact that the electron-cantilever interaction
is invariant in the transverse direction. The direct term from (\ref{b5}) is
not of any further interest, since it just renormalizes the equilibrium
position of the cantilever, and we assume that this correction is already
included in the definition of the distance $2a$. Due to the symmetry in the
problem the wave function $\chi _{\vec{k}}$ of the entire system including
the rigid square barrier factorizes as
\begin{equation}
\chi _{\vec{k}}(\vec{r})=\phi _{k_x}(x)\zeta _{k_{\parallel }}(\vec{\rho})
\label{b8.5}
\end{equation}
where $\rho $ labels the coordinates in the parallel directions in real
space and $\zeta _{k_{\parallel }}(\vec{\rho})$ is just a two dimensional
plane wave. We can use this to perform the frequency summation in (\ref{b6})
using the representation for $G^0$ (cf. Eq. (28) of Ref.\onlinecite{bi2}),
\begin{equation}
G^0(a,a;k_{\parallel ;}i\omega _n)=\sum_{k_x}\frac{\phi _{k_x}^{*}(a)\phi
_{k_x}(a)}{i\omega _n-\varepsilon _{\vec{k}}}\quad .  \label{b8}
\end{equation}
One further conceptual difficulty in calculating an exchange interaction
diagram is that an electron can, in principle, travel from the position of
the cantilever through the barrier to the fixed electrode, and back again.%
{\bf \ }This is why we have to take the propagators of the entire system as
indicated in (\ref{b5}) and (\ref{b6}). There is an analogous correction to
the left side propagator $\tilde{g}^0$, since an electron could travel to
the oscillator position on the right and back again. These processes would
allow the full semi-infinite propagators $g$ and $\tilde{g}$ to pick up a
dependence on the chemical potential on the left hand side and right hand
side, respectively. For the small barrier transparencies considered here,
contributions arising from these processes can be shown to be negligible
(see also sec. 3 of Ref.\onlinecite{bi2}). To a very good approximation we
can therefore replace $G$ by $g$ in (\ref{b5}) and (\ref{b6}). We also
replace in (\ref{b8}), the $k$ -sum representation for $g,$ the full
wavefunctions $\psi $ by the wave functions $\phi $, which are the analogous
solutions for the semi-infinite problem. These wavefunctions are given by
\begin{equation}
\phi_{k_x} (x)=\left\{
\begin{array}{ll}
{\displaystyle {4i{\tilde q} \over {\tilde D}}}
\sinh ({\tilde K}x)e^{-i{\tilde q}a} & 0<x<a \\
e^{-i{\tilde q}x}-e^{i{\tilde q}(x-2a)}%
{\displaystyle {2({\tilde K}\cosh {\tilde K}a+i{\tilde q}\sinh {\tilde K}a)
\over {\tilde D}}}
& a<x
\end{array}
\right.   \label{b9}
\end{equation}
where now ${\tilde{K}}=\sqrt{2mV_0-k_x^2}$, ${\tilde{q}}=\mid k_x\mid $ and $%
{\tilde{D}}=2({\tilde{K}}\cosh {\tilde{K}}a-i{\tilde{q}}\sinh {\tilde{K}}a).$
The $k$ -sum representation for $g$ is then
\begin{equation}
g^0(a,a;k_{\parallel };i\omega _n)=\sum_{k_x}\frac{4k_x^2\sinh ^2(\sqrt{%
2mV_0-k_x^2}a)}{2mV_0-k_x^2+2mV_0\sinh ^2(\sqrt{2mV_0-k_x^2}a)}\frac 1{%
i\omega _n-\varepsilon _{\vec{k}}}\quad .  \label{b10}
\end{equation}
In order to simplify the notation, we omit the dependence of our functions
on $k_{\parallel }$ and $a$ for the moment, and we set
\begin{equation}
f(k_x)=\frac{4k_x^2\sinh ^2(\sqrt{2mV_0-k_x^2}a)}{2mV_0-k_x^2+2mV_0\sinh ^2(%
\sqrt{2mV_0-k_x^2}a)}\qquad .  \label{b10.5}
\end{equation}
The Matsubara summation in (\ref{b6}) is standard and yields
\begin{equation}
\Sigma ^{1ex}(i\omega _n)=-(x_0V_0)^2%
\mathop{\displaystyle \sum }
_{k_x}f(k_x)\left[ \frac{n_B(\omega _c)+n_F(\varepsilon _{\vec{k}})}{i\omega
_n+\omega _c-\varepsilon _{\vec{k}}}+\frac{n_B(\omega _c)+1-n_F(\varepsilon
_{\vec{k}})}{i\omega _n-\omega _c-\varepsilon _{\vec{k}}}\right]
\label{b10.7}
\end{equation}
where $\varepsilon _{\vec{k}}=\frac{k^2}{2m}-\mu $ and $n_F$ and $n_B$ are
the Fermi and Bose distribution functions
\[
\begin{array}{ccc}
{\ n_F(\varepsilon )=\frac 1{e^{\beta \varepsilon }+1}} & \text{and} & {\
n_B(\omega )=\frac 1{e^{\beta \omega }-1}\qquad }.
\end{array}
\]
If (\ref{b10.7}) is considered at zero temperature the Fermi distribution
functions turn into $\theta $-functions and the Bose contributions vanish,
so that
\begin{equation}
\Sigma ^{1ex}(i\omega _n)=-(x_0V_0)^2\sum_{k_x}f(k_x)\left[ \frac{\theta
(-\varepsilon _{\vec{k}})}{i\omega _n-\varepsilon _{\vec{k}}+\omega _c}+%
\frac{\theta (\varepsilon _{\vec{k}})}{i\omega _n-\varepsilon _{\vec{k}%
}-\omega _c}\right] \quad .  \label{b11}
\end{equation}
As it turns out, the zero temperature approximation is not very accurate for
realistic parameters of a model system. The effects of residual temperature
will be considered more thoroughly in the discussion and the appendix.

The result of (\ref{b11}) can be analytically continued ( $i\omega
_n\rightarrow \omega +i\delta $ ) to yield the retarded function.

It should be remarked at this point that within the full Keldysh formalism
for fully non-equilibrium situations, the Dyson equation (\ref{b4}) in its
correct form would read
\begin{equation}
g^{+}=(1+g^r\Sigma ^r)g^{0+}(1+g^a\Sigma ^a)+g^r\Sigma ^{+}g^a  \label{b11.5}
\end{equation}
where the superscripts $r$ and $a$ refer to the advanced and retarded
functions respectively (cf. Eq. (5a) in Ref.\onlinecite{bi3}). Also the
retarded self-energy would not just be a convolution of $G^r$ and $D^r$, but
(cf. Eq. (6) in Ref.\onlinecite{bi3})
\begin{equation}
\Sigma ^r\propto (D^r*G^{-})+(D^{+}*G^r)\text{ \quad .}  \label{b11.7}
\end{equation}
However, in our quasi-equilibrium case both equations return to their usual
equilibrium form. The main point in which the CCNS formalism enters our
treatment is that we will renormalize the quantity $\gamma $ defined above
to obtain a correction for the current in the next section. It is well known
that the tunneling current across the unperturbed barrier is a non-linear
function of the applied bias $V$ for large enough biases, although the
expression can be linearized for small biases. The modulation of the barrier
width introduces a second, small energy scale, so that the resulting
contribution to the differential conductivity significantly depends on $V$
even for small values of the bias.

Following an approach by Rickayzen \cite{bi7} and Scalapino \cite{bi8} the $%
k $ -sum is assumed to have its strongest contributions coming from the
immediate vicinity of the Fermi surface. Going over to an integral
representation of the sum and changing the sign of $\varepsilon _{\vec{k}}$
in the first term of (\ref{b11}), we find
\begin{equation}
\Sigma _r^{1ex}(\omega )=-(x_0V_0)^2N^{1D}(0)f(\sqrt{2m\mu -k_{\parallel }^2}%
)\int_0^\infty \left[ \frac 1{\omega +\omega _c+\varepsilon +i\delta }+\frac
1{\omega -\omega _c-\varepsilon +i\delta }\right] d\varepsilon  \label{b12}
\end{equation}
where $N^{1D}(0)$ $=\frac{\sqrt{2m}}{2\pi }(\sqrt{\mu -\frac{k_{\parallel }^2%
}{2m}})^{-1}$ is the one dimensional density of states at the Fermi surface
including the sum over two spin directions.

At this point it would be natural to attempt to include the effects of
finite mechanical damping in the cantilever motion. However, it is difficult
to include such effects from first principles and beyond the scope of this
first approach prescribed herein. Instead, a simple way to model the effects
of damping is to change $\omega _c$ to the frequency for a damped classical
harmonic oscillator, satisfying the equation of motion
\begin{equation}
\ddot{x}+\frac{\omega _c}Q\dot{x}+\omega _c^2x=0  \label{b13}
\end{equation}
which has the solutions $\omega =\frac{i\omega _c}{2Q}\pm \omega _c\sqrt{1-%
\frac 1{4Q^2}}$. This replacement just introduces a finite imaginary part
into the Green's function to achieve line-broadening and finite peak heights
related to the $Q$ of a classical oscillator. For simplicity let $b\equiv
\omega _c\sqrt{1-\frac 1{4Q^2}}$ and $c\equiv \frac{\omega _c}{2Q}$ . The
self energy can then be written
\begin{equation}
\Sigma _r^{1ex}(\omega )=(x_0V_0)^2Nf\Sigma ^b(\omega )  \label{b14}
\end{equation}
where
\[
\Sigma ^b(\omega )=\int_0^\infty d\varepsilon \left[ \frac{(\omega
+b+\varepsilon )-ic}{(\omega +b+\varepsilon )^2+c^2}+\frac{(\omega
-b-\varepsilon )-ic}{(\omega -b-\varepsilon )^2+c^2}\right]
\]
\begin{equation}
=\frac 12\ln \left| \frac{(\omega -b)^2+c^2}{(\omega +b)^2+c^2}\right|
-i\left[ \pi -\arctan (\frac{b-\omega }c)-\arctan (\frac{b+\omega }c)\right]
\quad .  \label{b15}
\end{equation}
The real part of $\Sigma ^b(\omega )$ in (\ref{b15}) has the usual
logarithmic form for this kind of diagram and the imaginary part is a
smoothed-out version of the $\theta $ -functions occurring for zero damping.

\section{Corrections for the current}

\subsection{Derivation of the corrections}

The self energy obtained in the last subsection can now be inserted into the
simplified version of (\ref{b4}) to yield a correction for the right hand
side propagator $g,$ which is needed to calculate the corrections to the
current and its first and second derivatives. We will use the same
representation of the current as in Eq. (41) of Ref.\onlinecite{bi2} with
the only changes due to the three dimensional nature of our model, so that
the expression for the total specular current $I_x$ across the barrier is
proportional to the cross sectional area of the system. Dropping the
subscript, it reads
\begin{equation}
\langle I(V)\rangle =8e\nu ^{(2)}\int_{-\infty }^\infty \frac{d\tilde{\omega}%
}{2\pi }\int \frac{d^2k_{\parallel }}{(2\pi )^2}\frac{Im\gamma (0)Im\tilde{%
\gamma}(0)}{\mid \gamma (0)+\tilde{\gamma}(0)\mid ^2}\left[ n_F(\tilde{\omega%
}-\mu _L)-n_F(\tilde{\omega}-\mu _R)\right]  \label{c0.1}
\end{equation}
where $\tilde{\omega}=\omega _R+\mu _R=\omega _L+\mu _L$. For $T=0$ this can
be recast into
\begin{equation}
\langle I(V)\rangle =8e\nu ^{(2)}\int_{-\infty }^\infty \frac{d\tilde{\omega}%
}{2\pi }\int \frac{d^2k_{\parallel }}{(2\pi )^2}\frac{Im\gamma (0)Im\tilde{%
\gamma}(0)}{\mid \gamma (0)+\tilde{\gamma}(0)\mid ^2}\left[ \theta (\mu _L-%
\tilde{\omega})\theta (\tilde{\omega}-\mu _R)-\theta (\mu _R-\tilde{\omega}%
)\theta (\tilde{\omega}-\mu _L)\right] \quad .  \label{c0}
\end{equation}
In the limit $\mu _L\rightarrow \mu _R\rightarrow \mu $ (\ref{c0}) goes to
\begin{equation}
\langle I(V)\rangle =8e\nu ^{(2)}\int_\mu ^{\mu +eV}\frac{d\omega }{2\pi }%
\int \frac{d^2k_{\parallel }}{(2\pi )^2}\frac{Im\gamma (0)Im\tilde{\gamma}(0)%
}{\mid \gamma (0)+\tilde{\gamma}(0)\mid ^2}  \label{c1}
\end{equation}
(see the discussion in Ref.\onlinecite{bi2} for the derivation). In the
above expressions the imaginary parts of $\gamma $ still contain the $%
k_{\parallel }$ dependence, which is not indicated for simplicity. According
to its definition above, $\gamma $ on the right hand side will receive a
correction through the Dyson equation for $g$, yielding $\gamma =\gamma
^0+\gamma ^{\prime }$ , whereas for the left hand side there is no
correction, so that $\tilde{\gamma}=\tilde{\gamma}^0$. Following CCNS it is
assumed that for small bias $\gamma ^0\simeq \tilde{\gamma}^0$ holds well.
In order to get the first order correction to the integrand in (\ref{c1}) it
is expanded to first order in $\gamma ^{\prime },$ yielding
\begin{equation}
\frac{Im\gamma ^0(Im\gamma ^0+Im\gamma ^{\prime })}{\mid \gamma ^0+(\gamma
^0+\gamma ^{\prime })\mid ^2}\simeq \frac{(Im\gamma ^0)^2}{\mid 2\gamma
^0\mid ^2}+\frac{Im\gamma ^0Im\gamma ^{\prime }}{\mid 2\gamma ^0\mid ^2}+%
\frac{4(Im\gamma ^0)^2(Re\gamma ^0Re\gamma ^{\prime }+Im\gamma ^0Im\gamma
^{\prime })}{\mid 2\gamma ^0\mid ^4}\quad .  \label{c2}
\end{equation}
The expression for $\gamma ^0$ is already given in Ref.\onlinecite{bi2} and
can be written
\begin{equation}
\gamma ^0=K\frac{iq\cosh Ka-K\sinh Ka}{K\cosh Ka-iq\sinh Ka}=\frac{%
-(q^2+K^2)\cosh Ka\sinh Ka+iqK}{K^2\cosh ^2Ka+q^2\sinh ^2Ka}K\quad .
\label{c3}
\end{equation}
For arguments $Ka\sim 4$ or larger, which seems reasonable for real
tunneling processes, (\ref{c3}) goes asymptotically to the WKB-like result
\begin{equation}
\gamma ^0\simeq -K+\frac{4iqK^2}{2mV_0}e^{-2Ka}\quad .  \label{c4}
\end{equation}
This shows how the exponential dependence of the tunneling current on the
barrier width enters this calculation.

The correction $\gamma ^{\prime }$ is then obtained from the definition of $%
\gamma $ by differentiating the correction term in the Dyson equation
\begin{equation}
\gamma ^{\prime }=-\frac 1{2m}\lim_{x,x^{\prime }\rightarrow 0}\partial
_x\partial _{x^{\prime }}g^0(x,a)\Sigma (a,a)g^0(a,x^{\prime })=-\frac{8mK^2%
}{D^2}\Sigma (a,a)\quad ,  \label{c5}
\end{equation}
where again the dependence of $g$ on $k_{\parallel }$ is not indicated
explicitly. For reasons that will become apparent as we proceed, it is
convenient to calculate the dimensionless ratios $\frac{Im\gamma ^{^{\prime
}}}{Im\gamma ^0}$ and $\frac{Re\gamma ^{^{\prime }}}{Re\gamma ^0}$ . Written
in the following compact way we obtain
\begin{equation}
\left[ \frac{Im\gamma ^{\prime }}{Im\gamma ^0}\mid \frac{Re\gamma ^{\prime }%
}{Re\gamma ^0}\right] =\frac 8\pi x_0^2V_0(2m)^{\frac 52}\left( V_0-(\omega
+\mu -\frac{k_{\parallel }^2}{2m})\right) \sqrt{(\mu -\frac{k_{\parallel }^2%
}{2m})}\left[ \frac{Im(\frac{\Sigma ^b}{D^2})}{Im\gamma ^0}\mid \frac{Re(%
\frac{\Sigma ^b}{D^2})}{Re\gamma ^0}\right] \quad .  \label{c6}
\end{equation}

Using the expressions introduced for $\gamma ^0$ and $D$ , (\ref{c6}) then
yields
\begin{equation}
\left[ \frac{Im\gamma ^{\prime }}{Im\gamma ^0}\mid \frac{Re\gamma ^{\prime }%
}{Re\gamma ^0}\right] =C\left[ Re\Sigma ^b+Im\Sigma ^b\frac{K^2-q^2}{2qK}%
\mid 2\left( 2qKIm\Sigma ^b-(K^2-q^2)Re\Sigma ^b\right) \frac{e^{-2Ka}}{2mV_0%
}\right]  \label{c7}
\end{equation}
where $C=$ $\frac 8\pi mx_0^2\sqrt{\left( \mu -\frac{k_{\parallel }^2}{2m}%
\right) \left( V_0-(\omega +\mu -\frac{k_{\parallel }^2}{2m})\right) }$ .
Equation (\ref{c7}) clearly shows that the second part of the square bracket
is negligible for the small barrier transparencies considered here.

We can now conveniently write
\begin{equation}
Im\gamma ^{\prime }=Im\gamma ^0\eta  \label{c9}
\end{equation}
where we define $\eta =\frac{Im\gamma ^{\prime }}{Im\gamma ^0}$. Inserting (%
\ref{c9}) into (\ref{c2}) now shows that the third term in (\ref{c2})
arising from the expansion of the denominator will be exponentially smaller
than the first two, due to the exponential dependence of $Im\gamma ^0$ .
Thus (\ref{c2}) can be written
\[
\frac{Im\gamma ^0(Im\gamma ^0+Im\gamma ^{\prime })}{\mid \gamma ^0+(\gamma
^0+\gamma ^{\prime })\mid ^2}\simeq \frac{(Im\gamma ^0)^2}{\mid 2\gamma
^0\mid ^2}(1+\eta )\quad .
\]

So far all our results depend on $k_{\parallel },$ and as CCNS mention, the
unperturbed expression in the integrand of (\ref{c1}) is a strongly peaked
function around $k_{\parallel }=0,$ which mainly comes from its exponential
dependence on $Ka$ . More precisely, differentiating with respect to $V,$ (%
\ref{c1}) yields
\begin{equation}
\left\langle \frac{dI}{dV}\right\rangle =\int \frac{d^2k_{\parallel }}{(2\pi
)^2}h(eV+\mu -\frac{k_{\parallel }^2}{2m})\exp \left\{ -4a\sqrt{2m\left(
V_0-(eV+\mu -\frac{k_{\parallel }^2}{2m})\right) }\right\} \quad .
\label{c10}
\end{equation}
where $h$ is a function that varies only slowly over the range of the
integration compared to the exponential. Due to the strong exponential
behavior of the final term, the largest contribution to this integration
will come from the vicinity of the lower integration limit after
transforming to polar coordinates. The square root in the exponential in (%
\ref{c10}) can be expanded in $z=k_{\parallel }^2$ around $k_{\parallel }=0$%
. The integrand can then be cast into:
\begin{eqnarray}
\left\langle \frac{dI}{dV}\right\rangle &=&\int_0^\infty \frac{dz}{4\pi }%
h(eV+\mu -\frac z{2m})\exp \left\{ -4a\sqrt{2m\left[ V_0-(eV+\mu )\right] }%
\right\} \exp \left\{ \frac{-2az}{\sqrt{2m\left[ V_0-(eV+\mu )\right] }}%
\right\}  \label{c11} \\
\ &\simeq &\frac{\sqrt{2m\left[ V_0-(eV+\mu )\right] }}{8\pi a}h(eV+\mu
)\exp \left\{ -4a\sqrt{2m\left[ V_0-(eV+\mu )\right] }\right\}  \nonumber
\end{eqnarray}
where we have neglected the contribution coming from the upper limit.

\subsection{Discussion of $I(V),$ $dI/dV$ and $d^2I/dV^2$}

The current-voltage characteristic $I(V)$ can now be obtained by integrating
(\ref{c10}). For two reasons this procedure is not very rewarding. Firstly,
the integral is difficult to perform as the exchange term is no longer a
slowly varying function over the range of integration, and thus our previous
approximation of replacing $\int_x^{x+\delta x}f(x^{\prime })dx^{\prime }$
by $f(x)\delta x$ is not accurate. Secondly, the interesting corrections to $%
I(V),$ which come from the exchange diagram are already very small in $dI/dV$
(as will be shown later), so that the integration leading to $I(V)$ $\ $will
make them essentially undetectable. We will thus concentrate on the
expressions for the first and second derivatives of $I(V).$ The first
derivative $dI/dV$, i.e. the differential conductivity, is simply given by (%
\ref{c11}) and accordingly $d^2I/dV^2$ can be obtained by a further
differentiation. We thus have
\[
\frac{dI}{dV}=\frac{e^2\nu ^{(2)}}{2\pi ^2a}\sqrt{2m\left[ V_0-(eV+\mu
)\right] }\frac{(Im\gamma ^0)^2}{\mid 2\gamma ^0\mid ^2}(1+\eta )
\]
\begin{equation}
\frac{d^2I}{dV^2}=\frac{e^2\nu ^{(2)}}{2\pi ^2a}\frac d{dV}\left[ \sqrt{%
2m\left[ V_0-(eV+\mu )\right] }\frac{(Im\gamma ^0)^2}{\mid 2\gamma ^0\mid ^2}%
(1+\eta )\right] \quad .  \label{d1}
\end{equation}
As an approximation to $(Im\gamma ^0)^2/\mid 2\gamma ^0\mid ^2$ we have,
consistent with our approximation in (\ref{c10}), taken over the
WKB-expression (for $k_{\parallel }=0$) derived in Eq. (47) of Ref.%
\onlinecite{bi2} for further calculations,
\begin{equation}
\frac{(Im\gamma ^0)^2}{\mid 2\gamma ^0\mid ^2}=4\frac{(eV+\mu )\left[
V_0-(eV+\mu )\right] }{V_0^2}\exp \left\{ -4a\sqrt{2m\left[ V_0-(eV+\mu
)\right] }\right\} \quad .  \label{d11}
\end{equation}

We have used the expressions derived above to calculate the magnitude of
these effects on the I-V characteristics of a model system, consisting of a
single-crystal {\em Si} beam, rigidly clamped at both ends. The beam is
chosen to have a length of $500\AA $ and a cross-section of $100\AA \times
100\AA $, giving it a fundamental resonance frequency of $\omega _c/2\pi
=30GHz$ and a mass of $1.2\times 10^{-20}kg$. The effective spring constant
of the beam will be $420$ $N/m$. The parameters for the calculation are
listed in Table 1.
\[
\begin{array}{|l||l|l|ll}
\cline{1-4}
\text{Barrier} & V_0=5eV & \mu =2eV & a=5\AA & \multicolumn{1}{|l}{} \\
\hline
\text{Cantilever} & \omega _c=1.2\times 10^{-4}eV & m_c=1.2\times 10^{-20}kg
& k_c=420N/m & \multicolumn{1}{c|}{Q=10^3} \\ \hline
\end{array}
\]

{}{}It should be noted that these results are calculated at $T=0$, and that
a rather modest quality factor of $10^3$ for the oscillator has been chosen.
Mechanical oscillators at much lower $\omega _c$, but with $Q$'s of order $%
10^6$ or higher have been reported \cite{mr4}. Our choice of small $Q$ was
made to very roughly compensate for the finite temperature broadening
effects in any actual experiment. As described in the Appendix, a $Q$ of $%
10^3$ corresponds roughly to a measurement on this system at $1mK$, but with
a mechanical $Q$ possibly much higher than $10^3$.

Figures \ref{Fig5} and \ref{Fig6} show the first and second derivative of
the current density $j$ (in $A/m^2$). The unperturbed terms for $dI/dV$ and $%
d^2I/dV^2$ are not symmetric about the origin, since in our model we have
assumed that the Fermi energy is fixed on one side of the barrier, while it
is varied about this value by the bias on the other side. It can be seen
that the relative magnitude of the exchange contribution in $dI/dV$ to the
unperturbed value is about $10^{-7}$ if temperature effects are neglected
entirely. The form of the exchange part is similar to that observed for
localized phonons. Comparing these graphs, it is apparent that it would be
difficult to see the contribution of the exchange resonance after a final
integration to get $I(V)$. The most visible feature is the correction to $%
d^2I/dV^2$: the exchange correction here reaches a size comparable to the
unperturbed value, and the resonant form clearly distinguishes it from the
smooth background.

Since the graphs only give a qualitative impression of the peak strengths,
we now investigate them more thoroughly. The peaks in $dI/dV$ clearly result
only from contributions of the real part of the exchange term, cf. Eq. (\ref
{b15}). The term has poles at $\omega =\pm b=\pm \omega _c\sqrt{1-\frac 1{%
4Q^2}}$, which, in the bare term, leads to a peak height of $Re\Sigma ^b(\pm
b)=\mp \frac 12\left[ \ln 16+\ln Q^2\right] $, independent of $\omega _c$.
The origin of the peaks in $d^2I/dV^2$ is slightly more delicate. Here, the
largest contributions come from the terms including $\frac d{d\omega }\left[
Re\Sigma ^b(\omega )\right] $ and also $\frac d{d\omega }\left[ Im\Sigma
^b(\omega )\right] $. The first term has poles at
\[
\omega =\pm \sqrt{b^2+c^2\pm 2c\sqrt{b^2+c^2}}=\pm \omega _c\sqrt{1\pm \frac
1Q}\simeq \pm \omega _c(1\pm \frac 1{2Q})\quad ,
\]
where the choices of the signs are made independently. The peak heights
scale as $\pm Q/\omega _c$ for the outer (inner) poles respectively. The
second term has its poles exactly at $\omega =\pm b$ , with heights that
scale as $\pm 2Q/\omega _c$ , and width (FWHM) of $\omega _c/Q,$ and thus
the peaks in the second derivative do depend on $\omega _c$. To see how $Q$
quantitatively affects the peak heights, we have compiled the following
table of maxima of the peaks in absolute height occurring in the exchange
corrections to $dI/dV$ and $d^2I/dV^2$, where apart from $Q$ all other
parameters for the system remain unchanged.
\[
\begin{array}{|l||l|l|l|l|}
\hline
Q & 10^3 & 10^4 & 10^5 & 10^6 \\ \hline
dj/dV[\Omega ^{-1}m^{-2}] & 150 & 188 & 227 & 265 \\ \hline
d^2j/dV^2[G(\Omega V)^{-1}m^{-2}] & 0.23 & 2.1 & 21 & 210 \\ \hline
\end{array}
\]

Eventually, as the intrinsic $Q$ of the oscillator is increased, the
broadening at zero temperature will be determined by electronic damping of
the oscillator, given by higher order terms in the self energy. However for
the values of $Q$ in the table, we believe these effects are unimportant.

\section{Conclusion}

In conclusion we have investigated the first and most relevant self energy
corrections to the first and second derivatives of the $I(V)$ characteristic
in M-I-M tunneling due to the interaction with a single localized mechanical
mode of a movable tunnel junction. This represents the first step in
investigating the influence of, and back-action on, oscillatory mechanical
degrees of freedom in a mesoscopic tunneling system. We find that the
presence of this mechanical mode gives a pronounced signature in $dI/dV,$
and a much stronger one in $d^2I/dV^2,$ in the regions where the bias across
the junction equals the energy of the eigenmode of the oscillator. This
strong non-linear enhancement of the differential conductivity and its
derivative across the junction in this region can be viewed as arising from
the opening of a new phonon-assisted channel for inelastic tunnelling.

As the derivation of these results has been made using the general many body
formalism of CCNS, the approach may be readily extended. Possible
applications might include considering stronger bias to generate a fully
non-equilibrium situation, incorporating the band structure in the metals,
and elucidating the role of other many body interactions.

\section{Acknowledgements}

We gratefully acknowledge helpful discussions with Professor Sir Roger J.
Elliott at Oxford regarding theoretical specialties of this work. NFS
gratefully acknowledges financial support from Wolfson College, the
Department of Theoretical Physics and the Board of Graduate Studies of the
University of Oxford. Furthermore he would like to thank Ah San Wong for her
kind support in organizing the end phase of this project.

\section{Appendix}

\appendix

In this appendix we derive the range of validity of the zero temperature
calculation and establish qualitatively the changes to our results as finite
temperature effects start to compete with and finally dominate the intrinsic
losses of the mechanical oscillator. The first zero temperature
approximation which entered our calculation from before was in (\ref{b10.7})
for the Matsubara self energy.
\begin{equation}
\Sigma ^{1ex}(i\omega _n)=-(x_0V_0)^2\sum_{k_x}f(k_x)\left[ \frac{n_B(\omega
_c)+n_F(\varepsilon _{\vec{k}})}{i\omega _n+\omega _c-\varepsilon _{\vec{k}}}%
+\frac{n_B(\omega _c)+1-n_F(\varepsilon _{\vec{k}})}{i\omega _n-\omega
_c-\varepsilon _{\vec{k}}}\right] \quad .
\end{equation}
If one still assumes that the largest contributions to the $k_x$-sum come
from the vicinity of the Fermi surface one can approximate the sum similar
to (\ref{b12}) as
\[
\Sigma _r^{1ex}(\omega )=-(x_0V_0)^2N^{1D}(0)f(\sqrt{2m\mu -k_{\parallel }^2}%
)
\]
\begin{equation}
\times \left( n_B(\omega _c)\int_{-\infty }^\infty d\varepsilon \left[ \frac
1{\omega +\omega _c-\varepsilon +i\delta }+\frac 1{\omega -\omega
_c-\varepsilon +i\delta }\right] +\int_{-\infty }^\infty d\varepsilon \left[
\frac{n_F(\varepsilon )}{\omega +\omega _c-\varepsilon +i\delta }+\frac{%
1-n_F(\varepsilon )}{\omega -\omega _c-\varepsilon +i\delta }\right] \right)
\label{ap4}
\end{equation}

The two integrals can be considered separately and the first one gives
\begin{equation}
\int_{-\infty }^\infty d\varepsilon \left[ \frac 1{\omega +\omega
_c-\varepsilon +i\delta }+\frac 1{\omega -\omega _c-\varepsilon +i\delta }%
\right] =\wp \int_{-\infty }^\infty d\varepsilon \frac{2(\omega -\varepsilon
)}{(\omega -\varepsilon )^2-\omega _c^2}-2i\pi  \label{ap5}
\end{equation}
where $\wp $ denotes the principle value. The principle value integral is
seen to be zero, since the integral is antisymmetric. The imaginary part
turns out to be constant with respect to $\omega $. The second integral can
be recast into
\[
\int_{-\infty }^\infty d\varepsilon \left[ \frac{n_F(\varepsilon )}{\omega
+\omega _c-\varepsilon +i\delta }+\frac{1-n_F(\varepsilon )}{\omega -\omega
_c-\varepsilon +i\delta }\right]
\]
\begin{equation}
=\wp \int_{-\infty }^\infty \frac{d\varepsilon }{e^{-\beta \varepsilon }+1}%
\frac{2\omega }{\omega ^2-(\omega _c+\varepsilon )^2}-i\pi \int_{-\infty
}^\infty d\varepsilon \frac{\delta (\varepsilon +\omega +\omega _c)+\delta
(\varepsilon -\omega +\omega _c)}{e^{-\beta \varepsilon }+1}\quad .
\label{ap6}
\end{equation}
If that is compared to formula (\ref{b14}), one sees that the difference for
finite $T$ is that the integrals now get a smooth cut-off at $\varepsilon =0$
as opposed to the $\theta $-function cut-off from before. On introducing a
finite quality factor Q, (\ref{ap6}) goes over to
\begin{equation}
\int_{-\infty }^\infty \frac{d\varepsilon }{e^{-\beta \varepsilon }+1}\left[
\frac{(\omega +b+\varepsilon )-ic}{(\omega +b+\varepsilon )^2+c^2}+\frac{%
(\omega -b-\varepsilon )-ic}{(\omega -b-\varepsilon )^2+c^2}\right]
\label{ap6.5}
\end{equation}
in analogy to (\ref{b15}). We can thus choose either (\ref{ap6}) or (\ref
{ap6.5}) as a basis of reasoning, depending on the regime considered. As it
turns out, for the conceivable range of oscillator quality factors $Q$, the
linewidths induced by $Q$ will be very small compared to the temperature
broadening, even for temperatures as small as $\sim 1mK$. Thus, we assume
for now that $Q\rightarrow \infty $ and use formula (\ref{ap6}). The
difference in the behavior for the real part of (\ref{ap6}) above from its $%
T=0$ analog can be seen through partial integration:
\begin{equation}
\wp \int_{-\infty }^\infty \frac 1{e^{-\beta \varepsilon }+1}\frac{2\omega }{%
\omega ^2-(\omega _c+\varepsilon )^2}=-\frac \beta 4\int_{-\infty }^\infty
\frac{d\varepsilon }{\cosh ^2\frac \beta 2\varepsilon }\ln \left| \frac{%
\omega -\omega _c-\varepsilon }{\omega +\omega _c+\varepsilon }\right|
\label{ap7}
\end{equation}
where the integrated term has vanished at the boundaries. In the limit $%
\beta \rightarrow +\infty $
\begin{equation}
\frac \beta 4\frac 1{\cosh ^2\frac \beta 2\varepsilon }\rightarrow \delta
(\varepsilon )\quad .
\end{equation}
This way one can see how the $T=0$ result for the real part of (\ref{b14})
is recovered. The corresponding integration of the imaginary part gives:
\begin{equation}
-i\pi \left[ \frac 1{e^{\beta (\omega +\omega _c)}+1}+\frac 1{e^{-\beta
(\omega -\omega _c)}+1}\right]  \label{ap8}
\end{equation}
which is seen to have a smoothed out step behavior, being essentially
constant for $\mid \omega \mid \gg \omega _c$ and being more or less zero
for $\mid \omega \mid \ll \omega _c$ as is seen in the corresponding
expression in (\ref{b14}), with the only difference that the width of the
transition is now the width of the Fermi distribution, which is
approximately
\begin{equation}
3.5k_BT\simeq 3\times 10^{-4}T\;eV/K\quad .  \label{ap9}
\end{equation}
It is possible to determine phenomenologically what effect this will have
for the most ``observable'' of the peaks, i.e. the ones in $d^2I/dV^2$. The
peaks in that function are, as we found, determined by $d\Sigma
_r^{1ex}(\omega )/d\omega $. It can be seen directly from (\ref{ap8}) that
the derivative of the imaginary part, which gives the strongest
contributions to these features, exhibits a pair of Lorentzian peaks at the
resonance frequency of $\pm \omega _c$. At nonzero temperature these have a
width determined by the Fermi distribution, as shown in (\ref{ap9}). One can
also see the change in the behavior of the derivative of the real part of $%
\Sigma $ from (\ref{ap7}) without having to perform the integral.
Differentiating with respect to $\omega $ gives
\begin{equation}
-\frac \beta 4\int_{-\infty }^\infty \frac{d\varepsilon }{\cosh ^2\frac \beta
2\varepsilon }\frac{2(\omega _c+\varepsilon )}{\omega ^2-(\omega
_c+\varepsilon )^2}\quad .  \label{ap10}
\end{equation}
This can be viewed as being a sum of two convolutions of opposite sign in $%
\omega $ between the derivative of the Fermi function $1-n_F$ and the two
partial fractions of the second factor. The result of the convolution of
these two peaked functions is to a good approximation again a peaked
function, where the width of the resulting feature is the sum of the widths
of the convoluted components. Since for the moment we assume that the width
of our initial peaks (due to the quality factor $Q$) is practically
negligible compared to the thermal width of the Fermi function, the
effective width of the peaks will also be given by the thermal width. If, on
the other hand, in an intermediate regime (low $Q$, low $T$) the two effects
start to compete one would have to use (\ref{ap6.5}) for a correct
treatment. However, as mentioned above, the effect will just be that the
widths will add to give the result.

The second $T=0$ approximation was introduced in (\ref{c0}) for the current.
In analogy to the reasoning above, one can argue that if the Fermi factors
are kept the smoothed out cut-off will lead to similar effects, which will
enhance the total broadening by an additional factor of about two. In
summary, we can conclude that the contributions of the quality factor $Q$
and the temperature $T$ to the total linewidth will be about
\[
\begin{array}{lll}
\displaystyle{\left[ \frac{\delta \omega }{\omega _c}\right] _Q=\frac 1Q} &
\text{and} & {\ \displaystyle\left[ \frac{\delta \omega }{\omega _c}\right]
_T=6\times 10^{-4}\frac{eV}K\frac T{\omega _c}},
\end{array}
\]
and the total width will be approximately the sum of these two contributions.

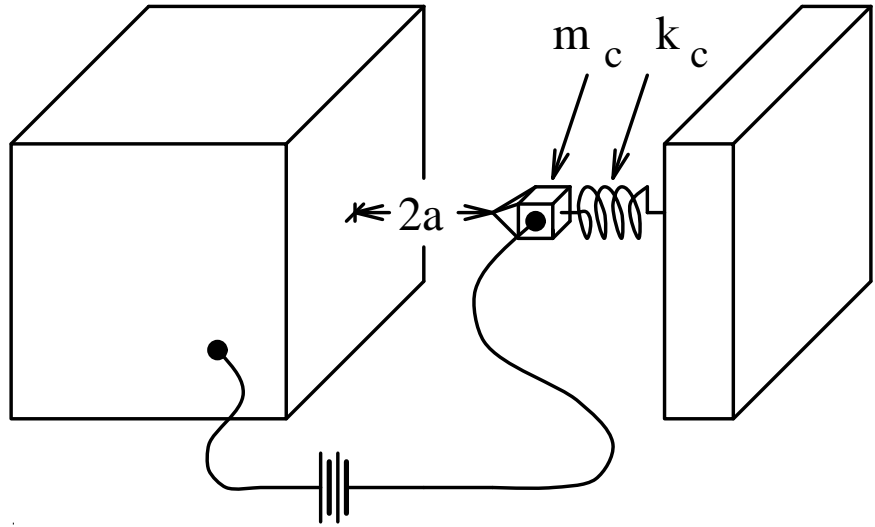
\begin{figure}[tbp]
\caption{Schematic view of tunneling electrode on cantilever with mass $m_c$
and spring constant $k_c$, placed a distance $2a$ from an infinitely massive
counterelectrode. The device is biased with a voltage $V$ and the resulting
current $I$ is measured.}
\label{Fig1}
\end{figure}

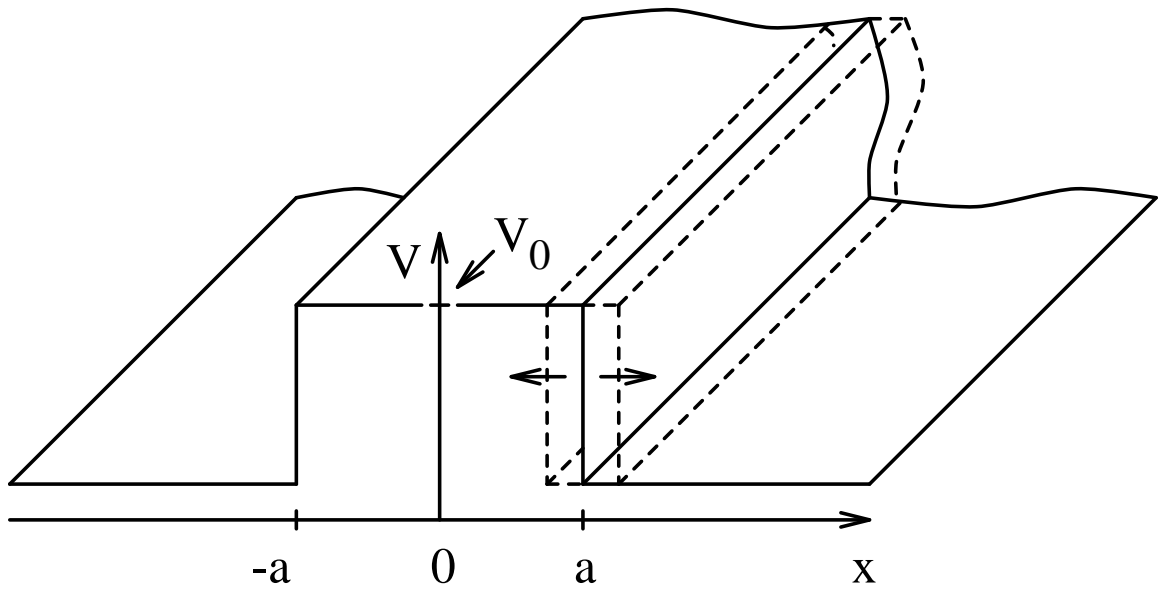
\begin{figure}[tbp]
\caption{Full potential barrier used in the calculation. The tunnel barrier
has height $V_0$ and equilibrium width $2a$. The right hand side of the barrier
corresponds to the position of the cantilever tip along the $x$-axis.}
\label{Fig2}
\end{figure}
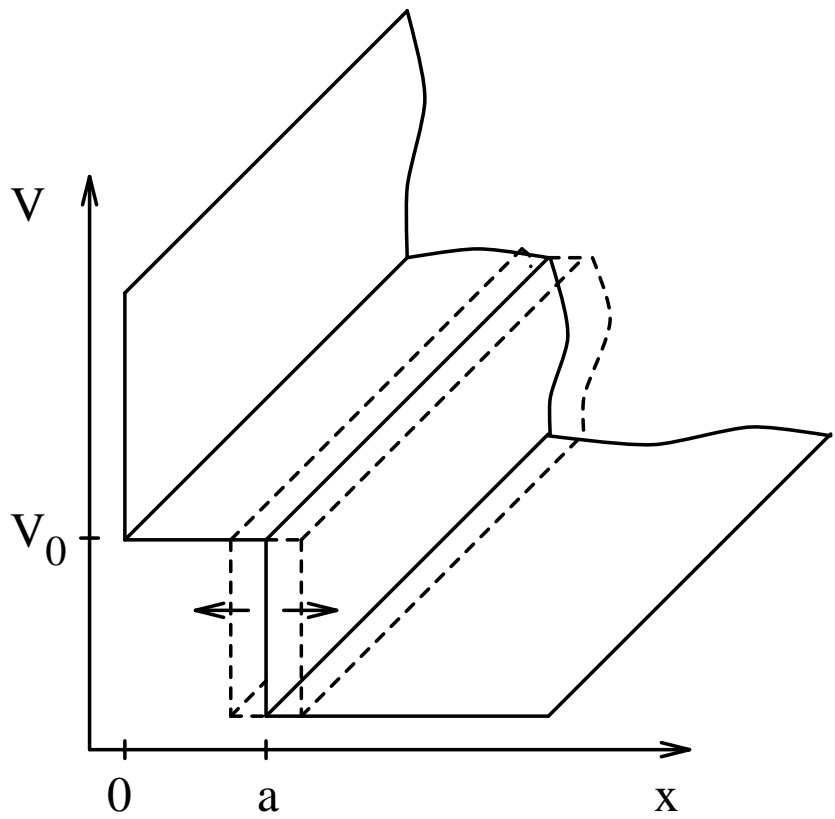
\begin{figure}[tbp]
\caption{Semi-infinite potential barrier used to split the Hamiltonian into
two pieces. The potential goes to infinity at $x=0$, is equal to $V_0$ for $%
0<x<a$, and is zero elsewhere.}
\label{Fig3}
\end{figure}
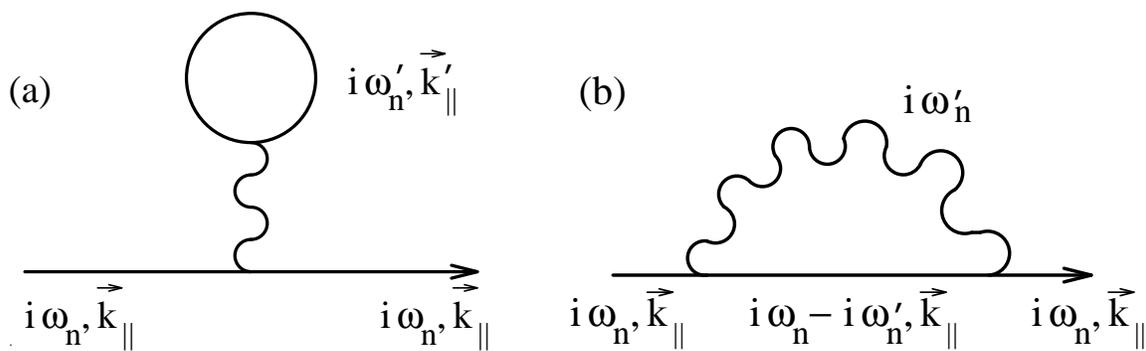
\begin{figure}[tbp]
\caption{Lowest order diagrams contributing to the self energy: (a) direct; and
(b) exchange. Electron paths are shown by straight lines and cantilever
``phonon'' paths are shown by sinuous lines.}
\label{Fig4}
\end{figure}
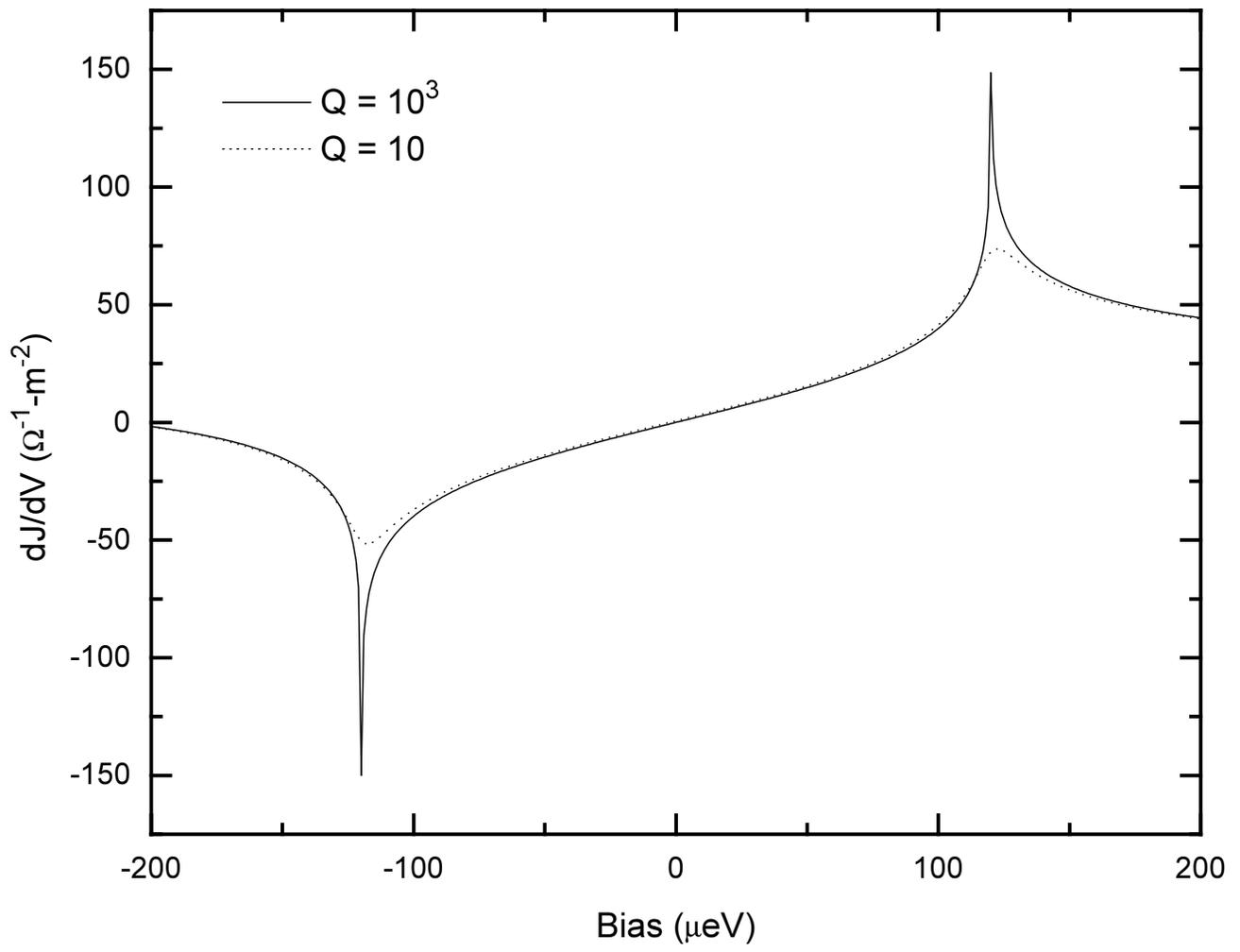
\begin{figure}[tbp]
\caption{Correction to the first derivative $dJ/dV$ of the current density:
the solid and dashed line are for cantilever quality factors $Q=10^3$ and $10
$ respectively.}
\label{Fig5}
\end{figure}
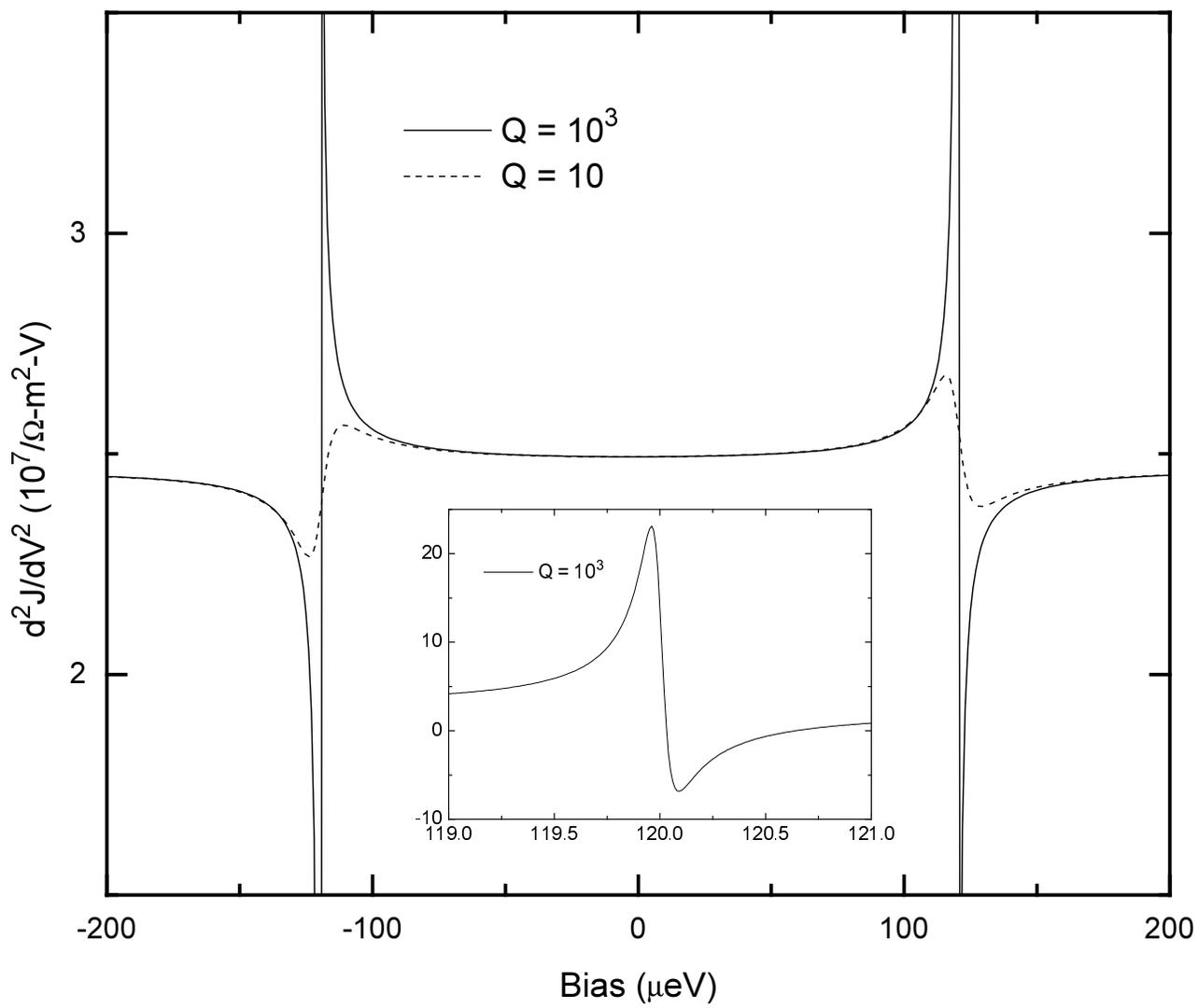
\begin{figure}[tbp]
\caption{Second derivative $d^2J/dV^2$ of the current density containing the
corrections considered in the text; the solid and dashed lines are for $%
Q=10^3$ and $10$ respectively. Inset: second derivative for $Q=10^3$, with
vertical scale sufficient to show full extent of peaks. (Axes have same
units as main plot).}
\label{Fig6}
\end{figure}

\end{document}